# Sociocultural Influences on Opinion Formation: Word of Mouth Dynamics, Mass Media and Behavioural Development

Elpida Tzafestas[0000-0001-5754-4146]
Laboratory of Cognitive Science, Department of History and Philosophy of Science, National and Kapodistrian University of Athens, University Campus, Ano Ilisia 15771, GREECE
etzafestas@phs.uoa.gr

**Abstract.** We study a society of agents belonging to a number of occupational/cultural groups that form opinions about others' situation in the same/different group. Opinions develop either by observation within one's own group or by directly interacting with members of other groups, therefore by word-of-mouth (WoM). Additionally, global mass media (MM) may be available that inform indirectly about the situation of the various groups. The sociocultural interplay of these processes and the degrees of relative exposure to each of the sources has diversified effects on final opinion and social attitude formation. In large and complex societies and groups where not everyone can physically interact by WoM with everyone about everything, these processes show potential for mass control and social automation engineering. Our model can also represent and be generally informative about segmented societies that consist of groups with different occupational and cultural characteristics and it can offer insights into social issues such as the generation gap, social cleavages and so on.

## 1 Introduction

People have opinions about almost everything, from potential solutions to everyday problems, such as which is the best route to work in times of congestion, to behavioural/emotional subjects, such as how one should respond to insult etc. They also have opinions on more theoretical moral and ideological issues that, although grounded in the real world, often have a religious or political flavour, such as whether same sex marriage should be allowed. People regularly express their opinions in various ways and they often change them, through discussions and debates with others, exposure to sources of information (books, media etc.), reflection etc.

Opinion update is a fundamental theme in the domain of social influence that is studied notably by sociologists and social psychologists [1-2]. Opinion modeling [3-7] is a computational research domain that offers many models and implementations, mainly related to opinion change/update and its emergent dynamics. The most widely used opinion update model is the bounded confidence model [5,7] where individuals may update their opinion in encounter with others if these are already close in opinion. Usual questions examined in opinion modeling are the formation of opinion groups and clusters [7-8], how agreement and consensus is formed [5,9], how polarisation and extremism emerge [10-13] etc. The opinions involved are typically of ideological nature, i.e. they can in principle be freely adopted, debated and updated.

In this paper, we adopt a slightly different perspective about opinion change and update. First of all, we study how individuals interact about issues that have a fairly objective value and if and how they may arrive to more-or-less the correct opinion: such issues are, for example, the condition of schools, the problems faced by nurses etc. In these cases, only the individuals related to school (e.g., as parents, teachers etc.) or to the hospital (especially nurses themselves) are knowledgeable and have a fairly objective first-hand view (moderated by individual personality, preferences, political orientation etc.). The rest of the individuals are inforrmed about these issues either by interaction with others (who have first-hand experience themselves or who have already acquired knowledge indirectly) or by getting information from newspapers, TV, books etc. Our goal is to study how close to reality are the opinions of the individuals, when both a process of direct interaction (word-of-mouth) exists and global/mass media are present. We are especially interested in seeing what happens when the global media transmit or project 'views' that are distorted in comparison to the real world and how and when and who can resist, make up for that divide and arrive to the right objective value of the issue. That question is of great significance to the study of spread of information and misinformation in society and is also a political theme related to propaganda, social manipulation, social automatism etc.

The rest of the paper is organised as follows. Section 2 presents the behavioural model with all the processes and the parameters involved. Section 3 presents the fundamental experiments and their results that show the basic tendencies emerging in this system, while section 4 studies a more complex environment with opinions about a large number of issues and the effect of global informers or 'leaders'. Section 5 investigates the dynamics of populations where two groups with different behavioural profiles coexist and whose opinions may diverge and clash. Next, section 6 examines what happens when the behavioural model itsef is not static but changes and develops over time. Finally, section 7 concludes and indicates some envisaged future work.

## 2 Behavioural Model

We have defined a social environment where a population of agents are interconnected in an interaction network and where a number of issues exist about which the agents may form opinions. Each issue has a social meaning that is assumed to concern a specific workplace and each agent is assigned to one of these workplaces. An agent spends some time initially in its workplace and then it may interact with one of the agents with whom it is connected or it may sit in front of a global medium (such as a TV channel, a news stream etc.) and be informed about any of the issues (including the one of its own workplace). We use the term 'word-of-mouth' in a generalised sense that includes both interactions face-to-face and interactions via a technical medium (via Facebook, by email etc.) The rest of the time, the agent is supposed to be idle or busy with activities that do not involve any of the social issues. Apart from the issue that concerns its own workplace, an agent is also interested in a few of the other issues and has an opinion about them that it updates through interaction with others (generalised word of mouth) and through obtaining information from global media. We assume that opinion update is spontaneous, that is, whenever exposed to some information, an agent updates a little its own opinion about the issue in question and does not have a threshold of confidence [5,7] that prevents getting closer to more distant opinions. Spontaneous imitation captures the fact that contact with and mere exposure to any information creates subconscious tendencies and biases, something that marketing and advertisement practices regularly exploit and that is investigated in psychology and communication studies [14-15].

We want to study how exposure to different sources of information (word of mouth or global media) may channel opinion development in different ways. We assume that the global media project random values about workplace issues that may be completely different from the actual objective values. On the contrary, we assume that all agents are sincere about their opinions during interaction with others and that they transmit their actual opinions, without any cheating. We then measure the average distance of an agent's opinions about its issues of interest from both their real/actual values and from the values projected by the media. Whenever the distance from real values is higher than that from the media projected values, we conclude that the agents have been manipulated by the external global media. This model allows us to identify the social environments and dynamics where the agents' own opinions are objective or where they are distorted by global media. Surely, in real life there are more processes involved, such as resistance to external information (whether information is acquired by word of mouth or by following global media). Still, we purport to show that exposure alone may already explain many things and may be decisive in many environments.

All the above are given in the general algorithm below:

*0. All issues (real values), agent opinions and media projected issue values are randomly initialised uniformly from 0 to 1.*

*1. At every simulation step, every agent interacts with one of the agents it is connected with. The agents execute in random order.*

*2. Agent execution:*

*a. With probability **pWork**, the agent goes to work and updates its opinion about the issue related to this workplace, as follows:*

*new opinion = old opinion + **rWork** * (perturbed real value – old opinion)*

*b. With probability **pOut**, the agent "goes out" to socialise and it selects one of the agents it connects with. If the two agents have a common issue then they interact. During interaction the initiating agent updates its opinion about the selected issue, as follows:*

*new opinion = old opinion + **rOut** * (perturbed partner opinion – old opinion)*

*c. If the agent has not gone out, then with probability **pTV**, the agent watches TV (or any global medium) about one of its issues of interest and updates its opinion about this issue, as follows:*

```
new opinion = old opinion + rTV * (perturbed projected media value – old
opinion)
d. Otherwise the agent remains idle, as far as opinion update is concerned
Perturbed (value) = value + dValue, dValue random in (-Noise,Noise)
```

Table 1 presents the behavioural parameters of the model. All agents are initialised with random workplace, probabilities and rates and they have 50% probability to be interested in any of the issues, thus on average they are interested in half of the issues in the environment and have a dynamic opinion about them. The interaction network between agents is also initialised randomly as either a random network, a small world network or a marketplace. The latter is not really a network but it is like an open space where every agent can meet every other agent, like in traditional markets.

**Table 1.** Behavioural parameters of the model

| | | | |
|---|---|---|---|
| **Number of agents** | N = 30, 50, 1000 | | |
| **Number of total issues (M)** | Default = 5 | **Number of issues per agent** | 50% of the issues |
| **Issues values (real, projected, opinions)** | From 0 to 1 | **Duration in steps** | 1000 |
| **Noise** | 0.02 | **Threshold (cf. section 6)** | From 0 to 1 (default=0.5) |
| **pWork** | From 0 to 1 | **rWork** | From 0 to 1 (default=0.1) |
| **pOut** | From 0 to 1 | **rOut** | From 0 to 1 (default=0.1) |
| **pTV** | From 0 to 1 | **rTV** | From 0 to 1 (default=0.1) |
| **pLeader (cf. section 4)** | From 0 to 1 | | |
| **Network types** | Random | K random connections per agent<br>N=30 (K=5,10,20), N=50 (K=5,15,25), N=1000 (K=5,50,1000) | |
| | Small world | Connections to 6 neighbours (3 before and 3 after) with probability 0.95 foreach connection and K random connections in the network<br>N=30 (K=30,70,120), N=50 (K=500,100,200), N=1000 (K=10,200,1000) | |
| | Marketplace | All to all connectivity | |

**Fig. 1.** (x: time, y: some issue values and distance) (left) Average distances from real values and projected values for a population of N=50 agents with 5 issues. (Right) Average distances

from real value and projected value of issue no. 1 for a population of N=50 agents with 20 issues. The right case converges much more slowly.

The duration of experiments (1000 steps) has been tuned to accommodate social environments of both faster or slower convergence. As Figure 1 shows, systems with relatively few agents and with a limited number of issues converge much faster than systems with a high number of issues. The same holds for very large populations.

## 3 Fundamental Experiments

All the experiments in this and the following sections are run for 1000 steps and all results are averages of 100 experiments. In all cases, we measure the average distance of the opinions of agents from the actual real values and from the values projected by the media. Because the latter are random, we would expect and would like the agents' average opinions to be closer to the real values rather than to the values projected by the media. In this section we show that this is not always the case.

**Table 2.** Comparative results for N=30 agents in two types of networks

| **K** | **Random network (30,K)** | | **K** | **Small world network (30, ±3, 0.95, K)** | |
|---|---|---|---|---|---|
| | **D(real)** | **D(projected)** | | **D(real)** | **D(projected)** |
| **pWork (0.8-1), High pOut (0.5-0.8), Low pTV (0.1-0.5)** | | | | | |
| **5** | .166 | .179 | **30** | .145 | .161 |
| **10** | .137 | .168 | **70** | .145 | .175 |
| **20** | .129 | .169 | **120** | .139 | .168 |
| **Marketplace** | | | | .104 | .239 |
| **pWork (0.8-1), Low pOut (0.1-0.4), High pTV (0.5-0.8)** | | | | | |
| **5** | .277 | .030 | **30** | .283 | .031 |
| **10** | .271 | .031 | **70** | .269 | .030 |
| **20** | .267 | .032 | **120** | .261 | .030 |
| **Marketplace** | | | | .223 | .127 |

Table 2 shows that when the probability of socialisation (pOut) is high and the probability of media following (pTV) low, then the average distance of the opinions of agents from the actual real values is less or equal than the average distance from the random values projected by the media. The opposite happens when pOut is low and pTV high: then the average distance from the real values is much higher than the distance from the projected values, thus the agents have a distorted view of reality and they tend to believe what the media diffuse. Note that there are practically no differences across the network connectivity (K) for random or small world networks, and very small differences between the two network types, but the marketplace environment allows overall closer distance to reality and higher distance to projected media for a combination of pOut, pTV values. This appears normal, since the marketplace environment supports interaction between any two agents in the population and thus allows information to spread fast in all directions.

Table 3 shows that in bigger populations the agents' opinions are overall closer to reality, given that we maintain the same total number of issues. Although the average connectivity in bigger networks is much lower than small ones (at most 100/1000=1/10 for N=1000 compared to 20/30=2/3 for N=30 in the case of random networks and similarly for the small world networks), the actual number of others with whom an agent interacts is in fact much higher, therefore the chance to get information closer to reality rises. This feature on the one hand yields smaller distance from reality when pOut is already high compared to pTV and on the other hand manages to moderate the higher distance from reality when pOut is low compared to pTV. As a general result, a higher volume of interaction and a higher number of others an agent meets allow the opinions to converge closer to reality.

**Table 3.** Results for big networks (N=1000 agents)

| **K** | **Random network (1000,K)** | | **K** | **Small world network (1000, ±3, 0.95, K)** | |
|---|---|---|---|---|---|
| | **D(real)** | **D(projected)** | | **D(real)** | **D(projected)** |
| **High pOut (0.5-0.8), Low pTV (0.1-0.5)** | | | | | |
| **5** | .107 | .231 | **100** | .114 | .161 |
| **50** | .097 | .234 | **200** | .112 | .175 |
| **100** | .099 | .243 | **1000** | .11 | .168 |
| **Marketplace** | | | | .092 | .225 |
| **Low pOut (0.1-0.4), High pTV (0.5-0.8)** | | | | | |
| **5** | .21 | .117 | **100** | .216 | .12 |
| **50** | .21 | .118 | **200** | .196 | .109 |
| **100** | .216 | .122 | **1000** | .223 | .124 |
| **Marketplace** | | | | .215 | .121 |

The same is shown in the results of Table 4 that are taken with a modified agent selection procedure (step 2b of the previous algorithm): the agent tries three times, and not just once, to find another agent in its network with a common issue. If for a third time it does not find one, only then does it abort its execution. With this setup, again the agents converge closer to reality.

Overall, we conclude that agents' opinions are closer to the real or the media distorted values according to the total volume of exposure and the diversification of exposure, i.e. the more time one spends socializing with more diversified and sincere sources (the other individuals in the society) the more objective its opinions (the closer to reality they are).

**Table 4.** Results for N=30 agents with enhanced partner choice function

| K | Random network (30,K) | | K | Small world network (30, ±3, 0.95, K) | |
|---|---|---|---|---|---|
| | D(real) | D(projected) | | D(real) | D(projected) |
| **High pOut (0.5-0.8), Low pTV (0.1-0.5)** | | | | | |
| **5** | .11 | .24 | **30** | .108 | .22 |
| **10** | .101 | .244 | **70** | .101 | .22 |
| **20** | .095 | .227 | **120** | .103 | .233 |
| **Marketplace** | | | | .097 | .237 |
| **Low pOut (0.1-0.4), High pTV (0.5-0.8)** | | | | | |
| **5** | .216 | .122 | **30** | .213 | .123 |
| **10** | .21 | .121 | **70** | .198 | .109 |
| **20** | .217 | .125 | **120** | .209 | .121 |
| **Marketplace** | | | | .213 | .125 |

## 4 Additional Processes

The results of the previous experiments showed that the volume of interaction with other individuals is the decisive factor for the final outcome in terms of matching of opinions to actual reality: when this volume is high the final opinions are closer to reality than to the values projected by the media but they are closer to media when this volume is lower than the volume of media following. These results were taken with a relatively low number of issues (5) of which only about half are of interest to any of the agents. What happens when the social world becomes more complex, in terms of number of issues? The agents may still be interested in about half of them, but do they have the chance to be fully informed about them through interaction with others? If not, then this would be an environment where they would be more easily manipulable by the global media. The results of Table 5 confirm that this is indeed the case. Even when the degree of socialisation is high compared to media following (pOut much greater than pTV), the average distance of agent opinions from reality is significantly greater than the corresponding distance from the media projected image.

One solution to this misinformation problem would be to introduce some cross-society “informers” whose role would be to transmit the true value of external issues to the agents so as to make up for the gap in person-to-person communication. Because the agents have first-hand information only about their work issue, the informers might be thought of as syndical leaders that connect workers of one type to the others. But this is a general property of the information setup and it can apply equally well to the generalised case of agents having first hand information about more issues, thus not necessarily only about their workplace. In the following experiment we examine a number of such “leaders” that visit the various work places and rotate randomly around them after every simulation step. Any leader is assumed to have a value for any issue that is the actual average of all workers in that issue (who have a first hand experience) perturbed by a degree of noise of up to 5%:

```
leader issue = avg. of workers value * dValue, with dValue a random in (0.95,1.05)
```

**Table 5.** Results for N=30 agents with large number of issues of interest (M=20)

| K | Random network (30,K) | | K | Small world network (30, ±3, 0.95, K) | |
|---|---|---|---|---|---|
| | **D(real)** | **D(projected)** | | **D(real)** | **D(projected)** |
| **High** pOut (0.5-0.8), **Low** pTV (0.1-0.5), **M=20** | | | | | |
| 5 | .212 | .156 | 30 | .213 | .154 |
| 10 | .215 | .155 | 70 | .216 | .153 |
| 20 | .202 | .151 | 120 | .210 | .153 |
| **Marketplace** | | | | .206 | .054 |
| **Low** pOut (0.1-0.4), **High** pTV (0.5-0.8), **M=20** | | | | | |
| 5 | .289 | .046 | 30 | .294 | .047 |
| 10 | .295 | .047 | 70 | .288 | .046 |
| 20 | .278 | .045 | 120 | .297 | .047 |
| **Marketplace** | | | | .291 | .047 |

Step 2a of the algorithm in presence of leaders then becomes:

*With probability **pWork**, the agent goes to work:*

*With probability **pLeader**, the agent meets the leader, if present in the work place, it gets informed about one of its issues of interest and updates its opinion about the issue related to this workplace, as usually:*

*new opinion = old opinion + **rOut** * (perturbed leader value – old opinion)*

*Otherwise, the agent updates its work issue opinion as before*

In the following experiments we measure the average distances for all the issues but the one concerning the workplace: the reason is that opinion about the workplace issue is always more objective because of continuous feedback at every step (step 2a of the algorithm), whereas the opinions about the ‘**other**’ issues are always less objective. So it is more accurate to evaluate the different behavioural options by studying the exact differences in this respect that are more significant than the regular averages of the previous sections. Table 6 presents a study on the number of such informers/leaders that are required to allow the agents to maintain opinions closer to reality than to media projected values. For the case of M=20, high pOut and low pTV, this value is around 4, that is at least 4 leaders with 50% influence are required to make all the population converge to opinions that are closer to reality than to what the media diffuse. However, when pOut is low and pTV is high, there is no number of leaders that could revert the tendency to believe the media rather the acquaintances, because even the theoretical maximum of 20 leaders (one per work place) cannot achieve this. For M=35 and high pOut, low pTV, it is shown that at least nine (9) leaders with 50% influence are required to achieve the same result and again no number of leaders works for low pOut and high pTV (results omitted for conciseness).

**Table 6.** Results for N=30 agents with large number of issues of interest and “syndical” leadership

| **L** | **Random network (30,5)** | | **Small world network (30, ±3, 0.95, K=5)** | |
|---|---|---|---|---|
| | **D(real.other)** | **D(proj.other)** | **D(real.other)** | **D(proj.other)** |
| **High** pOut (0.5-0.8), **Low** pTV (0.1-0.5), **M=20, Leaders (L), 50% influential** | | | | |
| 1 | .222 | .153 | .215 | .145 |
| 2 | .211 | .162 | .207 | .162 |
| 4 | .188 | .186 | .197 | .193 |
| 5 | .180 | .204 | .188 | .207 |
| 10 | .163 | .255 | .158 | .249 |

Similar results are obtained in media manipulation cases, where the complexity of the social environment hardly allows agents to obtain pertinent first and second hand information on several issues and therefore their opinions are often skewed away from reality and toward the values projected by the media. These cases include an implementation of social automatism, where media transmit values the farthest possible from reality (projected issue value = real value ± 0.5), and fake news, where the agent gets into contact with agents (or media) transmitting random values.

These results make us think that the complexity of the modern, urbanised and anonymous environments with very large populations, sparser interactions and many issues at stake make the individuals more vulnerable to external misinformation and manipulation. This is unlike more traditional environments with much smaller population sizes, fewer subjects of discussion and issues, much better knowledge of one another and limited presence of global media. Although propaganda has always existed in human history [16-17], its role and effect nowadays are much more pronounced.

# 5 Group Profiles

The society is not uniform in any dimension. This applies of course to the crucial parameters identified sofar, namely pOut and pTV. We can think of different groups having different setups for these parameters, as well as for the various network parameters. The people in northern countries (of the northern hemisphere) tend to socialise less than the people in southern countries and there are marked differences between, for example, Norway and Italy. These may be captured by the difference in pOut and pTV setups. Although geographic location, weather harshness etc. are not necessarily the only or the most important factors for socialisation, neither are they so for other cultural dimensions, as the debate on environmental or cultural or any other determinism has shown [18], our model allows some general conclusions to be made. Also, in the recent years with the advent and widespread adoption of social media, the landscape of social interaction has changed a little, and people may be going out less and interacting frequently with many more or unknown others via social media, or they may be interacting with social media on top of going out as before. These are all differences in cultural population behaviour that may be also captured by small variations of our model. In the next experiments, we are studying a population consisting of two groups with different pWork/pOut/pTV profiles that coexist and are interconnected. One crucial additional parameter that is investigated in the following experiments is the percentage of one group (the first) in the overall population; this is varied from 40% to 70%. Again, we measure distances for all the **'other'** issues (except the one concerning the workplace), especially since the various groups envisaged in what follows will often use different values for pWork.

In Table 7, we study two groups both with the usual high pWork (0.8-1), but the first highly socializing (high pOut and low pTV) and the second more media following (low pOut and high pTV). As is obvious from the table results, the first group's opinions are always closer to reality than to the values projected by the media, whereas for the second group the opposite is true. Moreover, the higher the share of the first group in the overall population, the closer their match with reality, while the second group's results are only a little affected, if at all. Because the two groups interact, the first group's opinions are dragged a little away from reality and the second group's opinions are dragged a little toward reality, compared to the corresponding cases of mono-group populations.

**Table 7.** Results in two social environments with two groups that differ in degree of going out versus following global media (see text)

| P(G1) | D(real.other) | D(proj.other) | D(real.other) | D(proj.other) |
|---|---|---|---|---|
| | | **Marketplace (30)** | | |
| 40% | .145 | .178 | .2 | .124 |
| 50% | .155 | .196 | .216 | .138 |
| 60% | .14 | .2 | .201 | .133 |
| 70% | .126 | .193 | .193 | .125 |
| | | **Small world network (30, ±3, 0.95, 70)** | | |
| 40% | .157 | .183 | .213 | .129 |
| 50% | .140 | .181 | .199 | .122 |
| 60% | .141 | .198 | .205 | .131 |
| 70% | .141 | .203 | .207 | .139 |

One important social behaviour difference, which is common in almost all of the western world, is that the older generations are generally socializing less than the younger ones and they are spending relatively more time in front of their TVs, due to both age fatigue and social isolation. In Table 8, we study two groups roughly corresponding to the youth and the retired people as described. The first population (the ‘youth’) have a high pWork, a high pOut and a low pTV. The second population (the ‘retired’ people) have a low pWork, a low pOut and a high pTV. Note that the low but nonzero pWork represents the fact that they maintain some occasional relation with their former workplace. The results obtained differ from these of Table 7. Now the low pWork of the ‘retired’ group makes them extremely vulnerable to the global media. If, on top of this, they have a high share in the population, they have a marked influence on the ‘youth’ agents that they interact with, so that the opinions of working and outgoing agents are still farther from reality compared to the values projected by the media. The results show that the share of the ‘youth’ in the population should be over 60%-70% for the youth to be closer to reality and that only by a little margin.

**Table 8.** Results in two social environments with two age groups (see text)

| P(G1) | D(real.other) | D(proj.other) | D(real.other) | D(proj.other) |
|---|---|---|---|---|
| | | **Random network (30,10)** | | |
| 40% | .211 | .157 | .318 | .045 |
| 50% | .183 | .16 | .301 | .051 |
| 60% | .161 | .163 | .271 | .052 |
| 70% | .151 | .174 | .265 | .058 |
| | | **Marketplace (N=30)** | | |
| 40% | .182 | .144 | .284 | .041 |
| 50% | .176 | .156 | .283 | .048 |
| 60% | .163 | .159 | .269 | .05 |
| 70% | .156 | .179 | .275 | .059 |

**Table 9.** Results in two social environments with two groups of young age that differ in employment status (see text)

| P(G1) | D(real.other) | D(proj.other) | D(real.other) | D(proj.other) |
|---|---|---|---|---|
| | | **Random network (30,5)** | | |
| 40% | .144 | .189 | .191 | .145 |
| 50% | .142 | .191 | .186 | .15 |
| 60% | .128 | .187 | .165 | .147 |
| 70% | .128 | .215 | .164 | .175 |

These results make us reflect about the case of two groups that differ in occupational condition instead of socialisation, like in youth unemployment. In Table 9, we study two groups that have both high pOut and low pTV, but the first one is regularly employed (pWork from 0 to 0.1), while the second one is only occasionally working and practically unemployed (pWork from 0 to 0.1). The fact that these second group agents have little first hand experience with a workplace (like the 'retired' people of the previous experiments) makes them again very vulnerable to any opinion. This shows in the table as higher distance from reality than from the opinions projected by the media. Because of high socialisation, their distorted opinions also spread in the whole population thus affecting the working group's opinions negatively and dragging them away from reality.

Finally, Table 10 gives the results of the same experiment as the two age groups experiment of Table 8, but this time with homophily. Homophily is the property of groups to tend to interact with ingroup rather than outgroup members. This is implemented in step 2b of the algorithm so that the agent tries three times to find another agent in its network which belongs to the same group. The results show that, with homophily, the first group is always closer to reality, even for low population share (below 50%), while the second group is always very close to what the media diffuse.

**Table 10.** Same as table 8 but with homophily (see text)

| P(G1) | D(real.other) | D(proj.other) | D(real.other) | D(proj.other) |
|---|---|---|---|---|
| | | **Random network (30,10)** | | |
| 40% | .146 | .149 | .269 | .028 |
| 50% | .145 | .176 | .291 | .037 |
| 60% | .135 | .203 | .293 | .044 |
| 70% | .118 | .212 | .283 | .05 |
| | | **Marketplace (N=30)** | | |
| 40% | .161 | .186 | .308 | .031 |
| 50% | .142 | .2 | .304 | .038 |
| 60% | .119 | .203 | .283 | .041 |
| 70% | .107 | .216 | .269 | .052 |

Once more, our results show that this kind of reasoning may be used as a basis to study and interpret several phenomena within the modern society, especially those that concern social segmentation and cleavages of all sorts: the generation gap, the dissonance between different occupational or economic classes, the clash between ethnic or religious groups and so on. In order to appease social tensions, then, one obvious political leverage means is to shuffle people of the various groups and nourish all kinds of contacts between them.

## 6 Behavioural Development

All the above results have been taken by assuming that all parameters, namely pOut and pTV, are static and do not change. In this section we relax this assumption and study what happens when behaviour changes and develops with time, as is the general rule with human behaviour. Firstly, in Table 11 we give for reference the results of the same two age groups experiment of Tables 8 and 10, but this with low pTV for the retired age too. This corresponds to the situation of the radio days or before the TV news-related programs became so many and so persistent as we see now. As the results show, both age groups are closer to reality than to the media projected values, with the retired group being farther than reality and closer to media than the young group, as expected.

**Table 11.** Same as table 8 but before the widespread presence of global media (see text)

| P(G1) | D(real.other) | D(proj.other) | D(real.other) | D(proj.other) |
|---|---|---|---|---|
| | | **Random network (30,10)** | | |
| 40% | .084 | .241 | .129 | .197 |
| 50% | .085 | .254 | .13 | .204 |
| 60% | .075 | .281 | .124 | .233 |
| 70% | .064 | .268 | .115 | .221 |
| | | **Marketplace (N=30)** | | |
| 40% | .09 | .246 | .133 | .199 |
| 50% | .081 | .249 | .127 | .204 |
| 60% | .07 | .27 | .116 | .223 |
| 70% | .061 | .275 | .108 | .231 |

Next, in Tables 12 and 13 we present the results of two experiments. In the first experiment, all agents may modify their pOut and pTV after every action. In the second one, the agents may modify accordingly their rOut and rTV. In the first case, the update is as follows (in steps 2b and 2c), while in the second case the update is exactly similar but it concerns rOut and rTV instead of pOut and pTV. The logic is simple: if a value is found close to the current opinion, the corresponding process (going out or TV following) is reinforced, otherwise it is attenuated. The threshold is tuned to 0.5, which is high but allows exploration of the opinion value space. Much smaller values up to 0.1 have been examined but they made no marked difference compared to the regular static case.

```
Step 2b:

diff = perturbed partner opinion – old opinion
new opinion = old opinion + rOut * diff
if (diff <= a threshold) pOut = pOut + 0.1 (until the maximum of 0.9)
else (diff > a threshold) pOut = pOut - 0.1 (until the minimum of 0.1)

Step 2c: similarly for pTV update
```

We may think of these experiments as experiments in trust development toward other agents or toward the media. Table 12 shows that the first development rule allows both groups to get closer to reality than to the image projected by the media, thanks to the second group developing higher pOut and lower pTV. On the contrary, Table 13 shows that both groups are closer to the projected media than to reality, even for high population shares for the first group. This is due to the fact that pOut and pTV are not affected by the update and both rOut and rTV rise for both groups. This shows once more that the crucial parameters are pOut and pTV and their relation, i.e., the crucial parameters are the ones controlling exposure to the various sources of information.

**Table 12.** Buildup of behaviour (probs) in the two age groups (see text)

| P(G1) | D(real) 1 | D(proj) 1 | POut 1 | PTV 1 | D(real) 2 | D(proj) 2 | POut 2 | PTV 2 |
|---|---|---|---|---|---|---|---|---|
| **Random network (30,10)** | | | | | | | | |
| 40% | .119 | .204 | .731 | .321 | .152 | .176 | .823 | .345 |
| 70% | .109 | .23 | .726 | .273 | .147 | .192 | .761 | .317 |
| **Marketplace (N=30)** | | | | | | | | |
| 60% | .116 | .214 | .74 | .282 | .149 | .189 | .823 | .311 |
| 70% | .101 | .229 | .73 | .26 | .134 | .198 | .77 | .286 |

**Table 13.** Buildup of trust (rates) in the two age groups (see text)

| P(G1) | D(real) 1 | D(proj) 1 | ROut 1 | RTV 1 | D(real) 2 | D(proj) 2 | ROut 2 | RTV 2 |
|---|---|---|---|---|---|---|---|---|
| **Random network (30,10)** | | | | | | | | |
| 40% | .272 | .076 | .79 | .788 | .329 | .021 | .872 | .896 |
| 70% | .232 | .111 | .741 | .669 | .32 | .031 | .785 | .889 |
| **Marketplace (N=30)** | | | | | | | | |
| 60% | .27 | .079 | .783 | .774 | .333 | .022 | .871 | .896 |
| 70% | .23 | .107 | .74 | .672 | .318 | .028 | .79 | .891 |

Apparently, one can experiment with other behaviour shaping options, for example direct imitation of pOut/rOut, which is a form of meta-imitation of the parameters that control opinion update (first-level imitation). Again, we know that in real life different sociocultural groups demonstrate different degrees of trust/distrust to the various sources of information, react differently to vast discrepancies between them and also oadapt differently to social change. This applies to very diversified types of groups, such as ethnic or religious groups, social or economic classes, age groups etc.

## 7 Conclusion

We have introduced a model of a population of agents that form opinions about issues with a fixed actual value and we have studied whether their opinions converge to these real values. Opinions are formed in two ways: either by directly interacting (word of mouth) with other agents or by interacting with global media sources. In both cases, opinion update happens always and is spontaneous, i.e., unlike bounded confidence models, it does not depend on a threshold. We have shown that if the exposure to global media is higher than the interaction with others, then the agents' opinions converge closer to the values projected by the media than to the real ones. All the parameters that affect exposure contribute to these differences. We have also shown that if the environment is overly complex with many issues at stake, the agents can not achieve convergence to reality in all of them. Measures such as introduction of informed leaders have only a limited applicability. Moreover, two social groups with different behavioural profiles may have totally different opinions, and this could explain phenomena such as the generation gap or some social clashes. Development of trust is also briefly examined.

This study is not only relevant to the investigation of the causes and consequences of misinformation, propaganda and the like, but also to the study of the emergence of clashes between groups with disjoint experiences and perceptions. Some immediate further studies are envisaged and concern groups with diversified access to the various social issues, tolerance to other opinions without the need to update oneself, dynamic rather than static social networks and, most importantly, cheating by individuals and influencers. More generally, the same kind of reasoning may be relevant for wider phenomena of emergence, channeling and stabilisation of cultural rules, norms and legislations as well as for the adoption of innovations.